\documentclass{article}
\usepackage{amsmath}
\usepackage{graphicx}
\usepackage{amssymb}
\usepackage{subfigure}

\newcommand{\beq}{\begin{equation}}
\newcommand{\eeq}{\end{equation}}
\newcommand{\bea}{\begin{eqnarray}}
\newcommand{\eea}{\end{eqnarray}}

\newcommand{\gsim}{\lower.7ex\hbox{$\;\stackrel{\textstyle>}{\sim}\;$}}
\newcommand{\lsim}{\lower.7ex\hbox{$\;\stackrel{\textstyle<}{\sim}\;$}}

\newcommand{\bi}{\begin{itemize}}
\newcommand{\ei}{\end{itemize}}

\usepackage{hyperref} 
\hypersetup{linktocpage=true}
\usepackage[all]{hypcap}

\begin{document}

\thispagestyle{empty}
\vspace*{-22mm}
\begin{flushright}
UND-HEP-10-BIG\hspace*{.08em}03\\
TUM-HEP-767/10\\
\end{flushright}
\vspace*{10mm}

\vspace*{10mm}

\begin{center}
{\Large {\bf\boldmath 
$D^0 \to \gamma \gamma$ and $D^0 \to \mu^+\mu^-$ Rates on an Unlikely Impact 
of the Littlest Higgs Model with T-Parity  }}
\vspace*{10mm}

{\bf Ayan\ Paul$^a$, Ikaros\,I.\ Bigi$^a$, Stefan\ Recksiegel$^b$} \\
\vspace{4mm}
{\small
$^a$ {\sl Department of Physics, University of Notre Dame du Lac}\\
{\sl Notre Dame, IN 46556, USA}\vspace{1mm}

$^b$ {\sl Physik Department, Technische Universit\"at M\"unchen,
D-85748 Garching, Germany}}

\vspace*{10mm}

{\bf Abstract}\vspace*{-1.5mm}\\
\end{center}
The decays $D^0 \to \gamma \gamma$, $\mu^+\mu^-$ are highly suppressed in the 
Standard Model (SM) with the lion's share of the rate coming from long distance 
dynamics; $D^0 \to \mu^+\mu^-$ is driven predominantly by 
$D^0 \to \gamma \gamma \to \mu^+\mu^-$. Their present experimental bounds are small, 
yet much larger than SM predictions. 
New Physics models like the 
Littlest Higgs models with T parity (LHT) can 
induce even large indirect CP violation in $D^0$ transitions. 
One would guess that LHT has a `fighting chance' to affect these $D^0 \to \gamma \gamma$, $\mu^+\mu^-$ 
rates in an observable way. We have found LHT contributions can be much larger than {\em short} distance 
SM amplitude by orders of magnitude. Yet those can barely compete with the {\em long} distance SM effects. 
{\em If} $D^0 \to \gamma \gamma$, $\mu^+\mu^-$  modes are observed at greatly enhanced rates, LHT scenarios 
will {\em not} be candidates for generating such signals. LHT-like frameworks will {\em not}  
yield larger $D^0 \to \gamma \gamma$/$\mu^+\mu^-$ rates as they are constrained by $B$ and $K$ rare decays.

\noindent

\newpage

\tableofcontents

\section{Introduction}

Compelling evidence for $D^0-\bar D^0$ oscillations has been presented  
 \cite{D0obs}. The interpretation of the oscillation 
parameters $x_D$ and $y_D$ inferred from the data has not been settled: while 
they could contain sizable contributions from New Physics (NP), they might still be 
compatible with what the SM can generate. Nevertheless it has sparked re-newed interest in building NP models that can affect 
$\Delta C =2$ dynamics significantly. This can be achieved even with models that had 
{\em not} been motivated by considerations of flavour dynamics. Littlest Higgs models with T parity provide an explicit class of examples that can generate sizable or even relatively large 
indirect CP violation in $D^0$ decays  \cite{DKdual}. There are also other scenarios 
for such novel effects  \cite{NIRETAL}. 

In LHT scenarios one gets new contributions also to $\Delta C =1$ decays without hadrons 
in the final state, namely $D^0 \to \gamma \gamma$, $\mu^+\mu^-$. Their rates are 
greatly suppressed both for fairly general reasons and those that are specific to SM dynamics. 
NP could then reveal its intervention through a significant enhancement in these rates. 
In this note we present a rather detailed analysis of the possible impact of LHT scenarios: 
in contrast to the situation with indirect CP violation in $D^0$ transition 
we do {\em not} find any {\em significant} enhancements from LHT dynamics. 

This note is organized as follows: after discussing short and long distance SM contributions to 
$D^0 \to \gamma \gamma$, $\mu^+\mu^-$ in Sect.\ref{SMPRED} we sketch LHT models 
as an interesting class of NP scenarios and their potential impact 
on $D^0 - \bar D^0$ oscillations in Sect.\ref{LHT}; then we analyze LHT contributions to 
$D^0 \to \gamma \gamma$, $\mu^+\mu^-$ and present our quantitative findings on their 
potential impact in Sect.\ref{RESULTS}; after our general comments about FCNCs in 
LHT-like frameworks in Sect.\ref{FCNC} we give our conclusions in Sect.\ref{CON}. 

\section{SM Contributions to $D^0 \to \gamma \gamma$, $\mu^+\mu^-$}
\label{SMPRED}

The rates for the modes $D^0 \to \gamma \gamma$, $\mu^+\mu^-$ are highly suppressed, 
since they must be driven by charm changing neutral currents and also require the annihilation of the 
$c$ and $\bar u$ quarks initially present in the $D^0$ meson;  $D^0 \to \mu^+\mu^-$ is further 
reduced greatly by helicity suppression. The question is how much they are suppressed, which dynamics drive them and whether they are short distance or long distance in nature. 

\subsection{$D^0 \to \gamma \gamma$}

There is an extensive literature both on $K_L \to \gamma \gamma$ and on the 
not yet observed $B^0 \to \gamma \gamma$. The former's reduced rate played an important 
role in the development of the SM, since it was one piece of evidence for Nature's suppression of 
strangeness changing neutral currents. It was also realized that $K_L \to \gamma \gamma$ is driven mainly by long distance dynamics. $B^0 \to \gamma \gamma$ on the other hand should 
be shaped mainly by short distance contributions. 

\begin{figure}[h!]
\begin{center} 
\includegraphics[width=11cm]{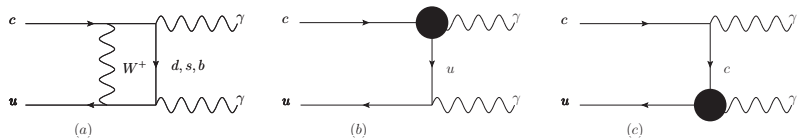} 
\caption{The diagrams contributing to $D^0 \rightarrow \gamma \gamma$. $(a)$ The 2PR (or 1PI) contribution. $(b)$$ \&$$(c)$ The 1PR contribution. The vertices stand for $c\rightarrow u \gamma$ diagrams} 
\label{fig:Dgg}
\end{center}
\end{figure}

$D^0 \to \gamma \gamma$ can be treated in general analogy to $B^0 \to \gamma \gamma$ with the amplitude described by two formfactors $A$ and $B$:
\beq
T(M\rightarrow \gamma\gamma)=\epsilon^\mu_1\epsilon^\nu_2\left[( q_{1\mu} q_{2\nu}-q_1 . q_2 g_{\mu\nu})A([Q\bar q])+i\epsilon_{\mu\nu\rho\sigma}q^\rho_1 q^\sigma_2 B([Q\bar q])\right]
\eeq
Those form factors receive contributions from two types of diagrams, the two-particle-reducible one (2PR) and 
the one-particle-reducible one (1PR) as shown in Fig.\ref{fig:Dgg}; the dark blob 
in the two diagrams on the right denote the effective $c\to u \gamma$ operator generated 
on the one-loop level. 
Both types of diagrams had been evaluated for $K_L \to \gamma \gamma$ in Ref. \cite{MaP} in 
terms of general quark masses. The pure electroweak contribution to the 1PR and 2PR amplitudes for $M^0 \equiv [Q\bar q] \to \gamma \gamma$ are given by 
\begin{eqnarray}
\nonumber A^{SD}([Q\bar q])&=&i\frac{\sqrt{2}G_F \alpha}{\pi}  f_M \sum_jV_{qj}^\ast V_{Qj}\left(A_j^{1PR}\right) \; , \\
\nonumber B^{SD}([Q\bar q])&=& \frac{\sqrt{2}G_F \alpha}{\pi}  f_M \sum_jV_{qj}^\ast V_{Qj}\left(A_j^{2PR}+A_j^{1PR}\right) \; , \\
A_j^{2PR}&=& (e_Q\pm 1)^2 \left[2+\frac{4 x_j}{x_M} \int_0^1\frac{dy}{y}\ln\left[ 1-y(1-y)\frac{x_M}{x_j}\right]\right]\nonumber \; ,\\
\nonumber A_j^{1PR}&=&\xi_M\{ e_Q(e_Q\pm 1)F_{21}(x_j)+e_Q F_{22}(x_j)\}\; , \\
F_{21}(x_j)&=&\frac{5}{3}+\frac{1-5x_j-2x_j^2}{(1-x_j)^3}-\frac{6x_j^2 }{(1-x_j)^4}\ln x_j\nonumber \; , \\
F_{22}(x_j)&=&\frac{4}{3}+\frac{11x_j^2-7x_j+2}{(1-x_j)^3}+\frac{6x_j^3}{(1-x_j)^4}\ln x_j\nonumber \; , \\
\xi_M&=&\frac{m_M^2}{16}\left<M^0\left| \frac{1}{p_1\centerdot q_1}+\frac{1}{p_1\centerdot q_2}+\frac{1}{p_2\centerdot q_1}+\frac{1}{p_2\centerdot q_2}\right| M^0\right> \; . 
\label{eqn:1PI}
\end{eqnarray}
We have used the following notation: $e_Q$ denotes the electric charge of the heavy quark $Q$, which 
is also carried by the lighter antiquark $\bar q$ inside the meson $M^0$; $x_M = m_M^2/m_W^2$ 
and $x_j = m_j^2/m_W^2$ where the nature of $j$ depends on $Q$: 
For $Q = s$ or $b$, the internal 
summation $j$ runs over the {\em up}-type quarks $u$, $c$ and $t$, while for $Q=c$ -- the case we will focus on -- 
$j$ runs over the {\em down}-type quarks $d$, $s$ and $b$. The $\pm$ in the 1PR and 2PR functions correspond to the cases of $Q=b,s$ and $Q=c$ respectively. The functions $F_{21}(x_j)$ and $F_{22}(x_j)$ together correspond to the $Q\gamma \bar q$ effective vertex for an on-shell photon and only differ from \cite{InamiLim} as they are valid for any arbitrary internal quark mass. $\xi_M$ is a hadronic factor that can be safely taken as one for the D meson as it should be in the nonrelativistic limit, which is a pretty good approximation for the D meson. 
$A$ and $B$ correspond to the final state photons being in a state of parallel and perpendicular polarization respectively. The branching fraction  and the CP asymmetry parameter $\delta$ is then given by\footnote{Removing the $^{SD}$ superscript from the expressions gives the general form to include all contributions.}
\begin{eqnarray}
{\rm BR}_{SD}(D^0\rightarrow \gamma \gamma)&=&\frac{m_{M}^3}{64 \pi}\left(\left|A^{SD}([Q\bar q])\right|^2+\left|B^{SD}([Q\bar q])\right|^2\right)\\
\delta&=&\frac{\left|A^{SD}([Q\bar q])\right|^2}{\left|A^{SD}([Q\bar q])\right|^2 + \left| B^{SD}([Q\bar q])\right|^2}
\end{eqnarray}
Due to the very different mass hierarchies 
for the up- and down-type quarks and the very peculiar structure of the CKM parameters $V_{ij}$, one finds that the 
same algebraic expression yields very different results for these radiative $K_L$, $D^0$ and $B^0$ modes. For the $K_L \to \gamma \gamma$ decay, the 2PR dominates over the 1PR contribution by a few orders of magnitude.
In $B^0 \to \gamma \gamma$ the 1PR contribution driven by $b \to s\gamma$ is comparable to the 2PR contribution  \cite{HK}. Even if  the $B^0 \to \gamma \gamma$ branching fraction is calculated solely from the $b \to s\gamma$ contribution, including the 2PR contribution raises the total branching fraction by about a factor of two and has been considered in quite a few works  \cite{HK,Bggother}.   

The situation is different for $D^0 \to \gamma \gamma$ -- and it is crafty in orders of QCD. The purely electroweak 
contributions from 1-loop without QCD are greatly dominated by 2PR over 1PR. Including QCD, leading logarithmic contributions of 1PR are 
significantly larger. Even more complete $\mathcal{O}(\alpha_S)$ corrections to the 1PR diagrams bring out the 
dominant contributions with amplitude $|A_{SD}(D^0 \to \gamma \gamma)| \simeq (2.35 \pm 0.50) \times 10^4 
\times |A_{SD}^{1-loop}(D^0 \to \gamma \gamma)|$   \cite{Greub,Burd}. From pure SD we get a branching fraction of: 
\beq
{\rm BR}_{SD}^{2-loop}(D^0\rightarrow \gamma \gamma)\simeq (3.6 - 8.1) \times 10^{-12}
\eeq
However, the $D^0\to\gamma\gamma$ transition is dominated by long distance effects  \cite{Ff,Burd}: 
\beq 
{\rm BR}_{SM}^{LD}(D^0\rightarrow \gamma \gamma)\sim (1 -  3) \times 10^{-8} \; . 
\eeq
This SM prediction is still substantially below the current experimental bound: 
\beq
{\rm BR}_{exp}(D^0\rightarrow \gamma \gamma)\sim 2.7 \times 10^{-5} \; . 
\eeq
The LD contribution calculated in  \cite{Burd,Ff} is model dependent and even though they give similar contributions to the branching fraction, they have disagreement in the phases of the amplitude and the relative magnitude of the CP even and CP odd amplitudes. Taking into consideration this uncertainty in the LD estimates, our estimate for the CP asymmetry parameter $\delta$ (using the amplitudes calculated in  \cite{Ff,Burd})  stands at
\begin{equation}
\delta\sim (0.95)0.5
\end{equation}

\subsection{$D^0 \to \mu^+ \mu^-$}

Realistically it seems one can improve the sensitivity for $D^0 \to \gamma \gamma$ only at an $e^+e^-$ machine like a Super-Flavour or a Super-Tau-Charm factory. The prospects for $D^0 \to \mu^+\mu^-$ are much better 
on one hand, since one has a fighting chance to probe it in hadronic collisions,  yet on the other hand the challenge 
is also much stiffer, since the rate for $D^0 \to \mu^+\mu^-$ suffers also from helicity suppression in the SM 
and most other NP scenarios. 

\begin{figure}[h!]
\begin{center} 
\includegraphics[width=8.5cm]{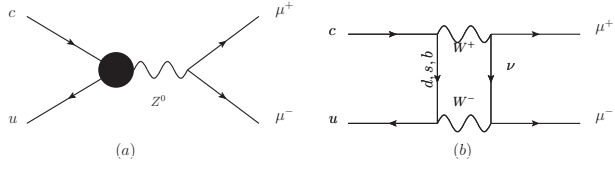} 
\caption{The diagrams contributing to $D^0 \rightarrow \mu \mu$. $(a)$ The $\bar{u}Z^0c$ effective vertex. $(b)$The $W^+W^-$ contribution.} 
\label{fig:Dmumu}
\end{center}
\end{figure}

The SM short distance contributions are given by the diagrams shown in Fig.\ref{fig:Dmumu}. 
Hence one obtains  \cite{BurasBmu}:
\begin{eqnarray}
B_{SM}^{SD}\left(D^0\rightarrow \mu^+\mu^-\right)&=&\frac{1}{\Gamma_D}\frac{G_F^2}{\pi}\left(\frac{\alpha}{4\pi\sin^2(\theta_W)}\right)^2f_D^2 m_\mu^2 m_D \sqrt{1-4\frac{m_\mu^2}{m_D^2}}\times\nonumber\\
&&\sum_{j=d,s,b}\left|V_{uj}^\ast V_{cj}\right|^2 \left(Y_0\left(x_j\right)+\frac{\alpha_s}{4\pi}Y_1\left(x_j\right)\right)^2\nonumber\\
Y_0\left(x_j\right)&=& \frac{x}{8}\left(\frac{x-4}{x-1}+\frac{3x}{(x-1)^2}\log(x)\right)\nonumber\\
Y_1\left(x_j\right)&=&\frac{4x_j+16x_j^2+4x_j^3}{3(1-x_j)^2}-\frac{4x_j-10x_j^2-x_j^3-x_j^4}{(1-x_j)^3}\ln x_j\nonumber\\
&&+\frac{2x_j-14x_j^2+x_j^3-x_j^4}{2(1-x_j)^3}\ln^2 x_j+\frac{2x_j+x_j^3}{(1-x_j)^2}L_2(1-x)\nonumber\\
&&+8x_j\frac{\partial Y_0{x}}{\partial x}\Bigg|_{x=x_j}\ln x_\mu\nonumber\\
L_2(1-x)&=&\int^x_1dt\frac{\ln t}{1-t}
\label{eqn:DmumuSM}
\end{eqnarray}
where $x_\mu=\mu^2/m_W^2$. Numerically one finds 
\beq
\rm BR_{SM}^{SD}\left(D^0\rightarrow \mu^+\mu^-\right)\sim 6\times10^{-19} \; , 
\eeq 
i.e., a hopelessly tiny number. 

However a less than a tiny prediction in SM $D^0 \to \mu^+ \mu^-$  can be made in analogy to $K_L \to \mu^+ \mu^-$: 
a $\gamma \gamma$ intermediate state contributes from  long 
distance dynamics  \cite{Burd}: 
\begin{equation}
B_{SM}^{LD}(D^0\rightarrow \mu\mu)\sim 2.7\times10^{-5} B(D^0\rightarrow \gamma\gamma) \; . 
\end{equation}
Hence one arrives at  
\beq
{\rm BR_{SM}}(D^0\rightarrow \mu\mu)\sim (2.7 - 8) \times 10^{-13} 
\eeq
using the SM estimate given above for $D^0 \to \gamma \gamma$.

\section{On LHT Scenarios}
\label{LHT}

The SM predictions presented above leave a large range in rates for these rare transitions, 
where NP could a priori make its presence felt. So-called Little Higgs models have been studied extensively over the past decade as a possible NP scenario  \cite{LH}. Rather than attempting to solve the hierarchy problem, they 
`delay the day of reckoning' and address a maybe secondary, yet very relevant problem, namely to reconcile the fact that the measured 
values of the electroweak parameters show no impact from NP even on the level of quantum corrections 
with the expectation that NP quanta exist with masses around the 1TeV scale so that they could be produced 
at the LHC. 

In this note we will analyze a subclass of Little Higgs models, namely 
Littlest Higgs Models with T parity (LHT).  In our view they possess several significant strong points: 
\begin{itemize}
\item 
They contain several states with masses that can be below 1 TeV; i.e. those states should be 
produced and observed at the LHC. 
\item 
Compared to SUSY models they introduce many fewer new entities and observable parameters. 
\item 
Their motivation as sketched above lies outside of flavour dynamics. Thus they have not been `cooked up' 
to induce striking effects in the decays of hadrons with strangeness, charm or beauty. 
\item 
Nevertheless they are {\em not} of the minimal flavour violating variety! 
\item 
Especially relevant for our study is the fact that they can have an observable impact on $D^0 - \bar D^0$ oscillations 
 \cite{DKdual,DMB}, as explained next. 
\end{itemize}
\subsection{Basics of LHT and Impact on $D^0 - \bar D^0$ Oscillations}

Little Higgs models in general contain a large {\em global} symmetry that gets broken spontaneously 
to a lower subgroup leading to the emergence of a set of scalar particles as Pseudo-Nambu-Goldstone Bosons (PNGB) of this broken symmetry that play the role of Higgs fields. These models push the Hierarchy Problem up to higher scales for its UV completion to deal with. To cancel out the radiative corrections to the SM Higgs mass one 
introduces  a new set of Gauge Bosons and new fermions with judiciously arranged gauge couplings; 
the quadratic mass renormalization to the SM Higgs mass is achieved with the help of quanta of the {\it same statistics} unlike in SUSY 
extensions of the SM.

These models have to address a major challenge: Since nothing could prevent the tree-level coupling of the SM particles to these new, mostly heavier, particles, amongst other things, the $\rho$ parameter gets shifted outside the 
allowed range for global symmetry breaking at the TeV scale. To address this, either the breaking scale had to be raised to above a few TeV or a new symmetry had to be incorporated into these models. 
Not surprisingly, preference has 
been given to keeping the breaking scale at a TeV so that new physics could be seen at the TeV scale.

Akin to what is generally done in SUSY, a discrete $\mathbb{Z}_2$ symmetry, T-Parity, has been postulated 
such that only pairs of the new quanta can couple to the SM states  \cite{CL1,CL2}. To accommodate this new symmetry into models that were already highly constrained, either an entire new set of scalars had to be brought into existence, or, as was done in the Littlest Higgs Model with T-Parity (LHT), a set of mirror fermions had to be postulated.\footnote{For a detailed description of the Littlest Higgs Model with T-Parity cf. \cite{LHTRev}.}

The symmetry structure of the LHT (which it inherits from the Littlest Higgs Model \cite{L2H}) is a global $SU(5)$ broken down to a global $SO(5)$ at the scale $f$. The T-parity is implemented through the CCWZ formalism 
 \cite{CL2,CCWZ} using non-linear representations of the symmetry group\footnote{There are other ways of implementing T-Parity in the Littlest Higgs Model in which only linear representations of the entire group are used. They typically involve starting with a larger global symmetry and expanding the gauge group or expanding the Higgs sector as done in the Minimal Moose Model \cite{Low}.}. The Higgs sector (both T-odd and even) is implemented as a nonlinear sigma model with a vacuum expectation value of $f$. The gauge group is a $[SU(2)\otimes U(1)]\otimes[SU(2)\otimes U(1)]$ broken down to $[SU(2)\otimes U(1)]_A$ which have the generators of the T-odd gauges and $[SU(2)\otimes U(1)]_V$ which become the SM electroweak gauge group. 

The particle content of the LHT stands as follows:\footnote{The QCD sector is strictly SM.}
\begin{itemize}
\item T Even
\begin{itemize}
\item All the SM particles.
\item A heavy partner to the SM top.
\end{itemize}
\item T Odd
\begin{itemize}
\item A set of T-odd $[SU(2)\otimes U(1)]$ heavy gauge bosons with the exact same couplings as the SM ones.
\item A set of T-odd heavy mirror fermions which are family-wise mass degenerate.
\item A heavy Higgs triplet and a singlet
\end{itemize}
\end{itemize}
While LHT have been crafted to deal with the non-observation of NP in the electroweak parameters even on the 
quantum level, they generate non-MFV dynamics. For imposing a $\mathbb{Z}_2$ symmetry in the LHT requires 
the introduction of the mirror fermions listed above. The two unitary $3\times 3$ matrices 
$V_{Hd}$ and $V_{Hu}$ describing the mixing of the 
up- and down-type mirror quarks to the down- and up-type quarks, respectively, have no reason to exhibit 
the same pattern as the CKM matrix. However, since the mirror quark matrices can be diagonalized 
simultaneously,  
the two matrices are related to each other by the CKM matrix \cite{VHDVHUCKM}: 
\begin{eqnarray}
V_{Hu}^{\dagger}V_{Hd}=V_{CKM}
\label{eqn:VVV}
\end{eqnarray}
Hence assuming some form for $V_{Hd}$ fixes $V_{Hu}$ and vice versa. Since the CKM matrix does not differ 
too much from the identity matrix, one realizes that LHT contributions exhibit 
a clear correlation of the phases in the charm and strange sector. 

The impact of LHT dynamics on $K$, $B$ and 
also $D$ transitions has been explored in considerable detail, and potentially sizable effects have  
been identified  \cite{angph,LHTBK,rareBuras, BlankeUpdate}. Among other things it was found that sizable indirect 
CP violation can arise in $D^0$ decays \cite{DKdual,DMB} very close to the present experimental upper bounds \cite{D0obs}. This realization 
then naturally leads to the question whether they could affect the modes $D^0 \to \gamma \gamma$, 
$\mu^+ \mu^-$ that are so highly suppressed in the SM in an observable way.

\subsection{LHT contributions to $D^0 \to \gamma \gamma$}

The LHT contributions to this decay channel will primarily come through the 1PR diagram where the $W$ boson will be replaced by its T-odd partner, the $W_H$ and the internal quarks will be replaced by their T-odd partner, the mirror quarks. These mirror quarks being very heavy ($\mathcal{O}(1 TeV)$), their contribution will be strictly short distance. The 2PR contribution benefits less from heavy fermion masses than the 1PR ones thus making it 
quite negligible. Redefining $x_{jH}=m_{jH}^2/m_{W_H}^2$, $x^\prime_H=a x_H$ with $j=d_H,s_H,b_H$, $x_{DH}=m_D^2/m_{W_H}^2$ and $V^{Hu}_{ij}$ as elements of $V_{Hu}$, where the subscript $H$ refers to the T-odd sector, Eq(\ref{eqn:1PI}) will be modified as follows:
\begin{eqnarray}
\nonumber A^{SD}_{LHT}([Q\bar q])&=&i\frac{\sqrt{2}G_F \alpha}{\pi}  f_M\sum_{j=d,s,b}\frac{v^2}{4f^2}\left[ V_{uj_H}^\ast V_{cj_H}\left(A_{jH}^{1PR}\right)\right]\\
\nonumber B^{SD}_{LHT}([Q\bar q])&=&\frac{\sqrt{2}G_F \alpha}{\pi}  f_M\sum_{j=d,s,b}\frac{v^2}{4f^2}\left[ V_{uj_H}^\ast V_{cj_H}\left(A_{jH}^{2PR}+A_{jH}^{1PR}\right)\right]\\
A_H^{2PR}&=& (e_Q\pm 1)^2 \left[2+\frac{4 x_H}{x_M} \int_0^1\frac{dy}{y}\ln\left[ 1-y(1-y)\frac{x_M}{x_H}\right]\right]\nonumber\\
\nonumber A_H^{1PR}&=&\xi_M e_Q\{\left((e_Q\pm 1)F_{21}(x_H)+ F_{22}(x_H)\right)-\frac{1}{6}F_{22}(x_H)-\frac{1}{30}F_{22}(x_H^\prime)\}\\
F_{21}(x_H)&=&\frac{5}{3}+\frac{1-5x_j-2x_j^2}{(1-x_j)^3}-\frac{6x_j^2 }{(1-x_j)^4}\ln x_j\nonumber\\
F_{22}(x_H)&=&\frac{4}{3}+\frac{11x_j^2-7x_j+2}{(1-x_j)^3}+\frac{6x_j^3}{(1-x_j)^4}\ln x_j\nonumber\\
\label{eqn:1PILHT}
\end{eqnarray}
where $v$ is the vacuum expectation value of the SM Higgs and $f$ is the vacuum expectation value which breaks the $SU(5)$ to $SO(5)$ in LHT. The two additional terms in the 1PR contribution which are proportional to $F_{22}(x_H)$ come from the effective $Q\gamma \bar q$ vertex with $Z_H$ or $A_H$  and heavy up type quarks in the loop. In Sect.\ref{RESULTS} we will combine these amplitudes with the SM ones.

\subsection{Impact on $D^0 \to \mu^+ \mu^-$}

The LHT contribution can come from three sources: 
\begin{itemize}
\item 
$Z_L$ penguins can contribute with the SM gauge boson in the loop being replaced by the corresponding heavy gauge boson and the internal SM quarks by their mirror partners. There can also be $Z_L$ penguins with only neutral gauge bosons as the $\bar{u}_lZ_Hu_H$ and $\bar{u}_lA_Hu_H$ vertices are possible. 
$Z_H$ or $A_H$ penguins are forbidden by T-Parity. 
\item 
There can be contributions from the box diagrams with the charged SM gauge bosons replaced by their T-odd partners and the same for the internal quarks. Box contributions can also come from the charged SM bosons being replaced by the neutral T-odd bosons and the internal quarks being replaced by the up-type mirror fermions
\footnote{For the Feynman diagrams cf.  \cite{rareBuras}}. 
\item 
Any of the aforementioned LHT contributions to $D^0 \to \gamma \gamma$ will affect also 
$D^0 \to \gamma \gamma \to \mu^+\mu^-$. 
\end{itemize}
Full amplitudes have of course to be gauge invariant. The $Z_L$ penguin contributions and those from the box diagrams have been calculated both in the unitary and 't Hooft-Feynman gauge \cite{rareBuras}.  The LHT contribution can be calculated by replacing the sum of $Y_0(x)$ and $Y_1(x)$ in Eq.\ref{eqn:DmumuSM} with 
$J^{\mu\bar{\mu}}(z,y)$ given by 
\begin{eqnarray}
J^{\mu\bar{\mu}}\left(z_i,y\right)&=&\frac{1}{64}\frac{v^2}{f^2}\Big[z_iS\left(z_i\right)+F^{\mu\bar{\mu}}\left(z_i,y;W_H\right)\nonumber\\
&&+4\left[G\left(z_i,y;Z_H\right)+G_1\left(z_i',y';A_H\right)+G_2\left(z_i,y;\eta\right)\right]\Big]
\end{eqnarray}
where
\begin{eqnarray}
&&z_i=\frac{m_{Hi}^2}{m_{W_H}^2}=\frac{m_{Hi}^2}{m_{Z_H}^2}, z'_i=az_i, a=\frac{5}{\tan^2\theta_W},y=\frac{m_{HL}^2}{m_{W_H}^2}=\frac{m_{HL}^2}{m_{Z_H}^2}\nonumber\\
&&y'=ay, \eta=\frac{1}{a}\nonumber\\
&&S\left(z_i\right)=\frac{z_i^2-2z_i+4}{(1-z_i)^2}\ln z_i+\frac{7-z_i}{2(1-z_i)}
\end{eqnarray}
The functions $F^{\mu\bar{\mu}}\left(z_i,y;W_H\right), G\left(z_i,y;Z_H\right), G_1\left(z_i',y';A_H\right), G_2\left(z_i,y;\eta\right)$ \cite{VHDVHUCKM} correspond to the contributions from the $W_HW_H, Z_HZ_H,A_HA_H,Z_HA_H$ box diagrams respectively. The function $S\left(z_i\right)$ is a contribution from the $Z_L$ penguin diagrams with internal mirror quarks. The replacement of a singularity \cite{rareBuras} in $J^{\mu\bar{\mu}}\left(z_i,y\right)$ with the function $S\left(z_i\right)$ was first pointed out by  \cite{Goto} and subsequently by  \cite{AgI} and incorporated as an update to FCNC calculations in $B$ and $K$ physics cited earlier \cite{BlankeUpdate}.

\section{Numerical Findings on LHT Contributions}
\label{RESULTS}

Before we go into the details of the LHT contributions, let us clarify the parameter space that was probed and the value of the LHT parameters that were kept fixed in this study. The LHT has 20 new parameters of which the ones which will be relevant to us are as follows:
\begin{itemize}
\item The LHT breaking scale $f=1TeV$  is fixed by choice.  
\item The masses of the three T-odd mirror quarks, $m_{dH}, m_{sH}, m_{bH}$ range from 300 to 1000 GeV. 
\item There are three independent mixing angles in 
$V_{Hu}$, $\theta_{12}^{Hu},\theta_{13}^{Hu},\theta_{23}^{Hu}$ and 
\item three irreducible phases in $V_{Hu}$, $\delta_{12}^{Hu},\delta_{13}^{Hu},\delta_{23}^{Hu}$. 
\end{itemize}
The parameter space used for these analyses is a set that satisfies all experimental constrains from $B$ and $K$ physics. A small parameter set was also used which did not follow such constraints to check whether constraints from $B$ and $K$ physics affects LHT contributions to $D$ physics.

\begin{figure}[h!]
\includegraphics[width=13cm]{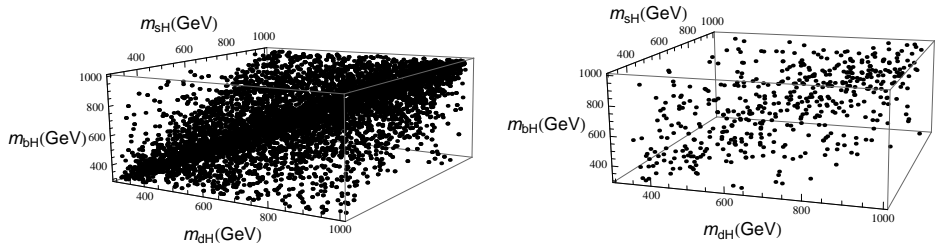}
\caption{Parameter space of the mass of the mirror quarks}
\label{fig:mass}
\end{figure}

\begin{figure}[h!]
\subfigure{
\includegraphics[width=13cm]{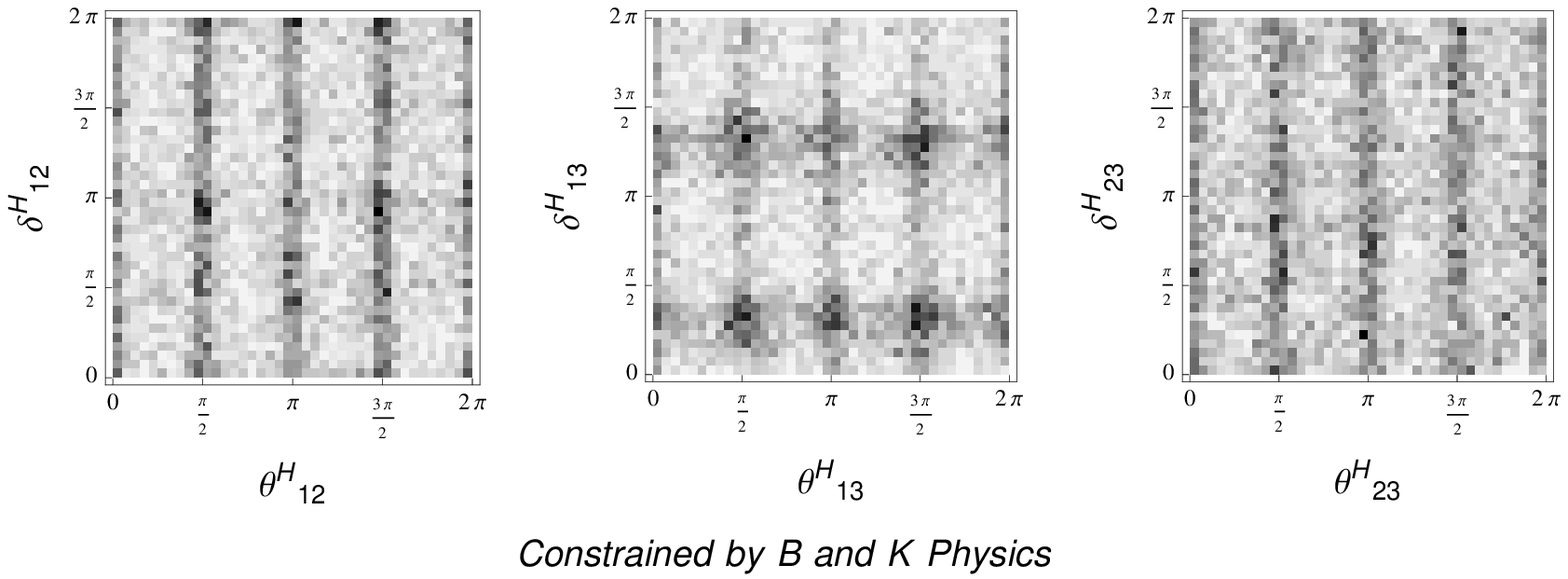}
}
\subfigure{
\includegraphics[width=13cm]{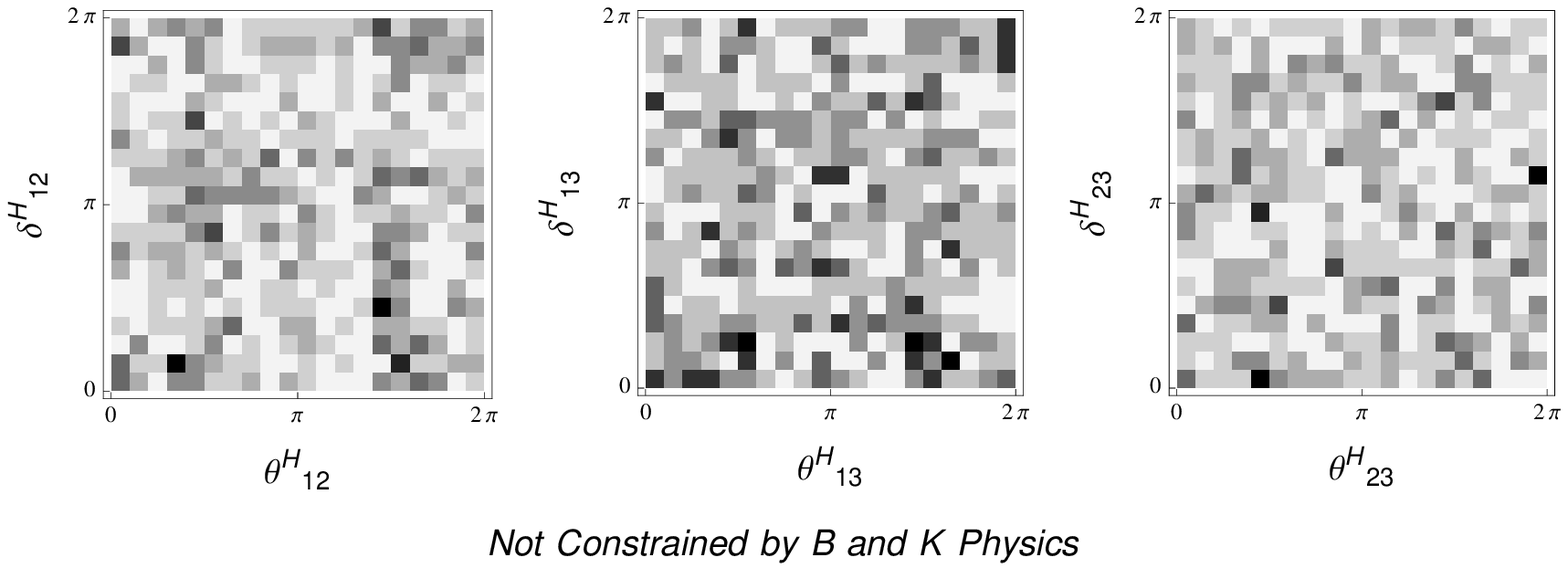}
}
\caption{Histogram of  the parameter space of the angles and phases in $V_{Hd}$. Counts in any bin are represented in grayscale, darker representing higher density }
\label{fig:angles}
\end{figure}

The mass spectrum for both the parameter sets is illustrated in Figs.\ref{fig:mass}. Using Eq(\ref{eqn:VVV}), the angles and phases of $V_{Hu}$ were calculated from those of $V_{Hd}$ and hence were constrained by $B$ and $K$ physics too for the first parameter set and not so for the second. Histograms of the parameter space of the angles and phases are shown in Figs.\ref{fig:angles}. The angles and phases are family-wise paired. 

\subsection{$D^0 \rightarrow \gamma \gamma$}
\begin{figure}[h!]
\includegraphics[width=13cm]{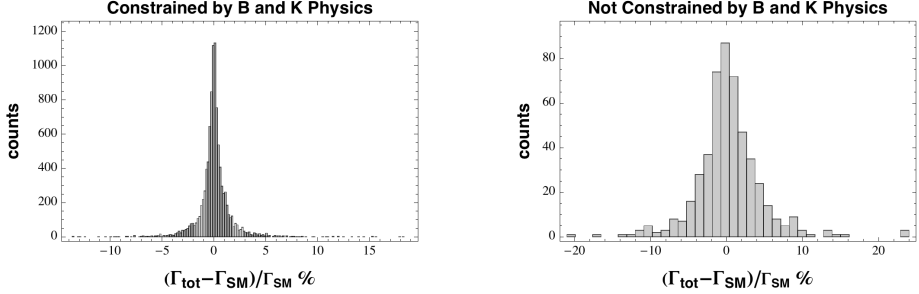}
\caption{Histogram of percentage enhancement to $\Gamma_{SD}(D^0\rightarrow \gamma\gamma)$ due to LHT contributions}
\label{fig:DggLHT}
\end{figure}

The LHT contribution to the branching fraction amounts at most to $O(10\%)$ of the SM short distance contribution; 
for most of the LHT parameter space it reaches merely a few percent as seen from Fig.\ref{fig:DggLHT}. The unconstrained parameter set gives us a very similar picture in Fig.\ref{fig:DggLHT}. The LHT contributions hardly affect the CP asymmetry parameter $\delta$. The dominance of LD contribution in the branching fraction and $\delta$ effectively swamps out any possible contribution that LHT can make to these. In view of the  
experimental challenges one can `realistically' hope to significantly improve the sensitivity 
for observing $D^0 \to \gamma \gamma$ only at a Super-Flavour Factory. Yet even if one managed to 
measure this transition one could never claim a case for having found LHT contribution considering the accuracy 
(or lack thereof) of the SM estimate given above.  

\subsection{$D^0 \rightarrow \mu^+ \mu^-$}
\begin{figure}[h!]
\subfigure[$m_{HL}=400GeV$]{\includegraphics[width=13cm]{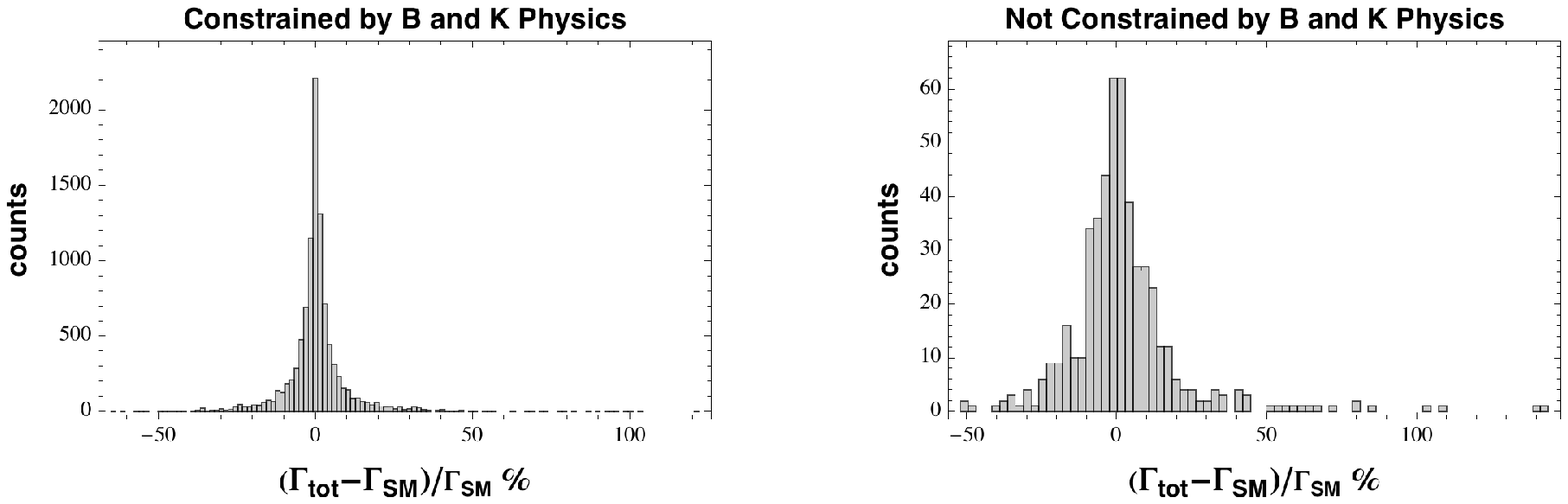}
\label{fig:Dmumu400}}
\subfigure[$m_{HL}=1100GeV$]{
\includegraphics[width=13cm]{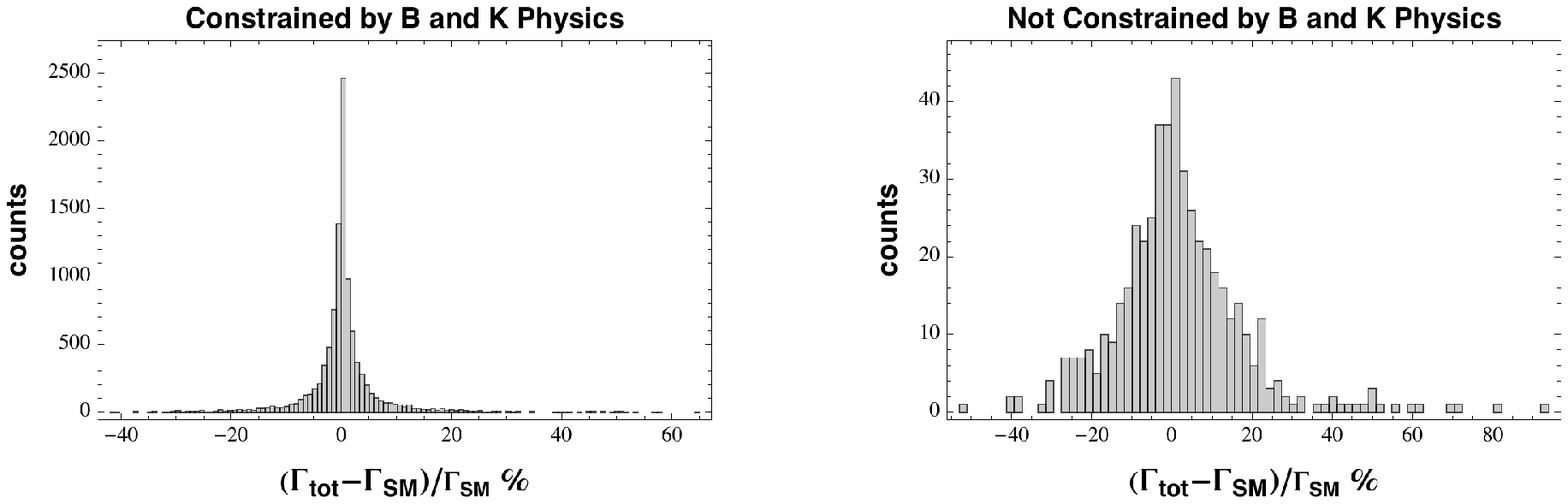}
\label{fig:Dmumu1100}
}
\caption[]{Histograms of  percentage enhancement to $\Gamma(D^0\rightarrow \mu^+\mu^-)$ due to LHT contributions for mirror neutrino mass of  \subref{fig:Dmumu400} $m_{HL}=400GeV$ and \subref{fig:Dmumu1100} $m_{HL}=1100GeV$}
\label{fig:DmumuLHT}
\end{figure}

The potential impact of LHT dynamics has been analyzed for different lepton masses, yet only data for two of the mirror neutrino masses will be represented by the graphs below. The first choice for the mirror neutrino mass is  400 GeV so that it falls within the mirror quark mass spectrum used in our studies. The second choice is a mass of 1100 GeV so that it lies outside the mirror quark mass spectrum.

As explained above, the dominant SM contribution to $D^0 \to \mu^+\mu^-$ arises from 
$D^0 \to \gamma \gamma \to \mu^+\mu^-$, where it hardly matters, whether the intermediate 
transition $D^0 \to \gamma \gamma$ is generated by long or short distance effects.

We see that LHT contributions are orders of magnitude larger than the SM short distance contribution to the extent that LHT contributions alone can be comparable to the long distance contribution to this channel in some regions of the parameter space. A very small region of the LHT phase space brings about six orders of magnitude enhancement over the SM SD contribution. However the SM LD contribution is projected to be six to seven orders of magnitude larger than the SM SD contribution and hence can easily be the dominant one. Enhancement to the total rate seems to be possible in rare cases, but only by a factor of 2 as seen in Fig.\ref{fig:DmumuLHT}. The constraints set by $B$ and $K$ physics do not change much of the analyses as is evident from Fig.\ref{fig:DmumuLHT}. However, it is almost impossible in this parameter space for LHT to provide the dominant contribution unless there is a larger mass splitting between the three generations of the mirror quark family.

\section{FCNCs in LHT-like frameworks}
\label{FCNC}

A careful analysis of the results of this study reveals that certain general conclusions can be drawn {\em beyond} the premises of LHT. In particular, the way with LHT affecting the rare decay channels depends purely on the 
{\em flavour} structure of LHT and {\em not} on the way this NP model is implemented as a {\em whole}. 
What defines the flavour structure of such models are:
\begin{itemize}
\item A second sector of fermions that are an exact copy of the SM ones.
\item  Mass mixing matrices which are unitary and loosely connected to $V_{CKM}$ (Eq.\ref{eqn:VVV}).
\item Possible large angles and phases in the mass mixing matrices.
\item Possible large hierarchies in the masses of the mirror quarks.
\item A symmetry, like T-Parity, segregating the NP sector from the SM sector, hence forbidding tree level FCNC.
 \end{itemize}
We have seen that LHT can generate a significant effect in the $D^0-\bar{D}^0$ oscillations  \cite{DKdual}. However, we see that is {\em not} true for the $D^0\to \gamma\gamma$ and $D^0\to\mu^+\mu^-$ channels. The reasons are as follows: Both the $D^0\to \gamma\gamma$ and $D^0\to\mu^+\mu^-$ channels are dominated by SM LD contributions which are larger than the SM SD contributions by orders of magnitude. Since tree level coupling to the heavier gauge bosons are forbidden by T-parity, the only LHT contributions are through loops, which are essentially SM SD operators because of the flavour structure of LHT. It is true that LHT with its heavier gauge bosons and heavy quarks can produce orders of magnitude {\em enhancements} to SM {\em SD} contributions -- but that typically falls shy or 
at best equals the LD contributions that these channels get from SM LD operators.

In   \cite{Gol} an interesting point has been shown that if NP can make significant contribution to $D^0-\bar{D}^0$ oscillations, then it can enhance the $D^0\to\mu^+\mu^-$ channel well beyond the SM. We do not 
disagree in general. Yet our study shows that LHT and LHT-like  \cite{CL2,Low}  frameworks cannot produce a significant contribution to
$D^0\to\mu^+\mu^-$ rate beyond the SM while a significant or even dominant signal can occur for 
$D^0-\bar{D}^0$ oscillations. When a NP has `construction plans' not only for the charm sector, but also 
for strange and beauty sectors -- what one has for a LHT-like framework -- there are connections for
charm, strange and beauty hadrons. Weak experimental constraints are not very stringent in $D$, but are 
very impressive in $B$ and $K$ physics. Hence these latter constraints can and have been used extensively to constrain the parameter space of any NP Models. 

In our studies we have compared $D^0 \to \gamma \gamma$, $\mu^+\mu^-$ with $K_L \to \gamma \gamma$, 
$\mu^+\mu^-$ and $B^0 \to \gamma \gamma$, $\mu^+\mu^-$. From these numerical calculations, we can conclusively prove that given the constraints from $B$ and $K$ physics, significant effects over and above SM are possible in  $D^0-\bar{D}^0$ oscillations, but not in $D^0\to \gamma\gamma$ and $D^0\to\mu^+\mu^-$ decays. 

The reason for this is as follows. LHT gives us the freedom of choosing large mixing angles and phases in the mixing matrices and also in the mass hierarchies of the quarks. Hence, large effects over SM SD, and possibly over SM LD, can be observed if we utilize both these freedoms. However, experimental constraints from $B$ and $K$ physics force us to choose between either large angles and phases or large mass hierarchies. In the current study we have chosen to live with large angles and phases rather than large mirror quark mass hierarchies. On the other hand, we could have made the mixing matrices ($V_{Hd}, V_{Hu}$) very diagonal and had large mass hierarchies but that would imply possibilities of the existence of quarks beyond the current experimental reach. In either case, LHT makes significant contributions over SM SD rates but fails to overpower SM LD rates. Moreover, if any NP model is able to make a large impact on the $D^0\to \gamma\gamma$ decay, it will also, indirectly, help in increasing the SM contribution to $D^0\to\mu^+\mu^-$ as that depends primarily on the two photon unitary contribution \cite{Burd}. This might well wash out any NP contribution to the $D^0\to\mu^+\mu^-$ decay. Hence, in the absence of large mass hierarchies amongst the new set of quarks, it is not possible to generate large effects in the $\Delta F=1$ processes although the same can be done in the $\Delta F=2$ processes, even if we allow for large angles and phases. In the absence of large hierarchies in the new quark sector, unitarity of the mass mixing matrices result in very tiny FCNCs, something akin to what is seen in FCNCs in the SM. The possibility of such large mass hierarchies are limited by experimental limits on $B$ and $K$ decay branching fractions and CP violating parameters if we chose to use large angles and phases in the mixing matrices. Making the other choice does not help either. Hence experimental observation of large effects in $\Delta F=1$ processes will automatically lead to a loss of viability of models of this nature.

The Littlest Higgs Model with T-Parity has undergone extensive scrutiny in the past few years as a major candidate for viable Little Higgs class of theories. Many suggestions have given for theoretical restructuring and for avoiding heavy constraints from experimental bounds \cite{Low,Hill,Papa}. The beauty of our analysis is that it is immune to changes in the way the specific flavour of the Littlest Higgs Model with T-Parity is implemented; it only depends on  the final flavour structure of the fermionic sector, which remains unchanged in all these models. Hence, our conclusions are more general than the specific model that we have worked with.

Recently some light has been shed on what could possibly be UV completion of the effective Littlest Higgs Model \cite{Yavin,Csaki,Freitas}. The primary motivation for these models is the cancellation of anomalies \cite{Hill,Hill2} that arise from the Wess-Zumino-Witten \cite{WZ,Witten} terms in the Lagrangian, which can save the lightest T-odd particle as a dark matter candidate. Furthermore, they address the obvious problem of the hierarchy between the 10 TeV and the Planck Scale. Some of these models can possibly introduce new TeV scale particles into the low energy effective theory which, if they have the correct quantum numbers, can bring about new contributions to FCNCs. A more careful look at these brings us to some more general conclusions. UV completion models for the Little Higgs Models are usually constructed with the following constraints in mind which are interconnected amongst themselves

\begin{itemize}
\item The breaking scale of the effective Little Higgs Models is preferred to be around 1 TeV.
\item FCNCs do not suffer from contributions significant enough to break the $\rho$ parameter, i.e., the tree level contribution to FCNC from NP is naturally suppressed at such low breaking scales.
\item Enhancement of FCNCs can appear only through loop contributions.
\end{itemize}

Under such conditions, any new contributions can at best be the size of the ones we have seen from LHT. Hence, we reemphasize, the conclusions that we have drawn shall hold good. On a final note we would also like to comment that since the lightest T-odd particle as a dark matter candidate need not be absolutely stable (\`a la proton), T-Parity (or any other discrete symmetry protecting it) need not be exact. How inexact the discrete symmetry can be is, of course, an open question and depend largely on how it is implemented.
\section{Conclusions}
\label{CON}

With the SM predicting tiny rates for $D^0 \to \gamma \gamma$ and $D^0 \to \mu^+\mu^-$ observing 
those modes would reveal the intervention of NP. That statement is still valid. What we have found in this 
note is that LHT dynamics could not provide significantly enhanced rates even for those scenarios 
of LHT parameters that can generate observable indirect CP violation in $D^0$ transitions. To be sure LHT 
contributions can greatly enhance SM short distance rates even by orders of magnitude -- in particular for 
$D^0 \to \mu^+\mu^-$ -- yet the SM short distance amplitude is so tiny relative to their long distance counterparts 
-- at least as they are presently estimated -- that the total rate is increased only very moderately. In that sense our findings are negative, though 
still significant: while LHT dynamics 
can generate striking effects in $D^0 - \bar D^0$ oscillations, they can barely enhance the rates for 
$D^0 \to \gamma \gamma$ and $D^0 \to \mu^+\mu^-$ beyond what one might conceivably predict for 
SM long distance contributions.

\section{Acknowledgements}
\label{ACK}

We wish to thank Monika Blanke for her helpful comments on this work. This work was supported by the NSF under the Grant No. PHY-0807959.

\end{document}